\documentclass[referee]{chjaa}          

\usepackage{graphicx,times}             
\input{epsf.sty}                        
\input{psfig.sty}                       

\begin{document}

\title{Simultaneous multi-wavelength observations of the TeV Blazar Mrk 421
during February $-$ March 2003: X-ray and NIR correlated variability}

   \volnopage{Vol.0 (200x) No.0, 000--000}      
   \setcounter{page}{1}           

   \author{Alok C. Gupta
      \inst{1}\mailto{}
   \and B. S. Acharya
      \inst{2}
   \and Debanjan Bose
      \inst{2}
   \and Varsha R. Chitnis
      \inst{2}
   \and Jun-Hui Fan
      \inst{1}
      }

   \institute{Center for Astrophysics, Guangzhou University, Guangzhou 510006, China \\
             \email{acgupta30@gmail.com}
        \and
             Tata Institute of Fundamental Research, Homi Bhabha Road,
             Colaba, Mumbai - 400 005, India \\
          }

   \date{Received~~2007 month day; accepted~~2007~~month day}

   \abstract{
In the present paper, we have reported the result of simultaneous
multi-wavelength observations of the TeV blazar Mrk 421 during February
$-$ March 2003. In this period, we have observed Mrk 421 using Pachmarhi
Array of \v Cerenkov Telescopes (PACT) of Tata Institute of Fundamental
Research at Pachmarhi, India. Other simultaneous data were taken from
the published literature and public data archives. We have analyzed the
high quality X-ray (2-20 keV) observations from the NASA Rossi X-Ray
Timing Explorer (RXTE). We have seen a possible correlated variability
between X-ray and J band (1.25 $\mu$) near infrared (NIR) wavelength.
This is the first case of X-ray and NIR correlated variability in Mrk 421 or
any high energy peaked (HBL) blazar. The correlated variability reported
here is indicating a similar origin for NIR and X-ray emission. The
emission is not affected much by the environment of the surrounding
medium around the central engine of the Mrk 421. The observations are
consistent with the shock-in-jet model for the emission of radiations.
   }

   \authorrunning{Gupta et al.}            
   \titlerunning{Variability of Mrk 421}  

   \maketitle

%
%
\section{Introduction}           
\label{sect:intro}

A small subgroup of radio-loud active galactic nuclei (AGNs) show
significant flux variability in the complete electromagnetic (EM)
spectrum, variable polarization and their radiation at all wavelengths
is predominantly non-thermal. They are known as blazars, which is a
collective name of subclasses (BL Lac objects, optically violent
variables OVVs, high polarization quasars HPQs and flat spectrum
radio quasars FSRQs) of radio-loud AGNs.  On a unified model of
radio-loud AGNs based on the angle between the line of sight and
the emitted jet from the source, blazars jet make angle of
$<$ 10$^{\circ}$ from the line of sight (Urry \& Padovani 1995).
Since, blazars emit radiation
in the complete EM spectrum which gives an excellent opportunity
to study the spectral energy distribution (SEDs). It is found from
observations that blazars SEDs have two peaks. The first component
peaks any where from IR to optical in so called red blazars or low
energy blazars (LBLs) or radio selected blazars (RBLs) and at UV/X-ray
in so called blue blazars or high energy blazars (HBLs) or X-ray
selected blazars (XBLs). Its origin is synchrotron emission from high
energy electrons in the jet. The second component extends up to
$\gamma-$rays, peaking at GeV energies in RBLs and at TeV in XBLs.
The electromagnetic emission is dominated by synchrotron component
at low-energy and at high-energy by inverse Compton component
(Coppi 1999, Sikora et al. 2001, Krawczynski 2004).

Mrk 421 is the nearest detected TeV BL Lac object (redshift z = 0.031).
It was first noted to be an object with blue excess which later turned
out to be an elliptical galaxy with bright point like nucleus (Ulrich et
al. 1975).  Since the energy of synchrotron peak of the source is higher
than 0.1 keV, it is classified as a HBL. Mrk 421 was the first extragalactic
object discovered at TeV energies (Punch et al. 1992).  This source was
later confirmed by the {\it high energy gamma ray astronomy} (HEGRA) group
(Petry et al. 1996). It is also one of the TeV blazars detected by
{\it energetic gamma ray experiment telescope} (EGRET) instrument in
the 30 MeV - 30 GeV energy range by the {\it Compton gamma ray observatory}
(CGRO) (Thompson et al. 1995). This source has been detected by the other
detectors like, the {\it imaging Compton telescope} (COMPTEL) on board
CGRO at the 3.2 $\sigma$ level in the 10-30 MeV energy range
(Collmar et al. 1999) and the {\it solar tower atmosphere \v Cerenkov effect
experiment} (STACEE) in the 140 GeV energy band (Boone et al. 2002).

Mrk 421 variability has been studied in all EM regimes in isolation. An
exhaustive compilation of radio data at 22 and 37 GHz, spanning for about
25 years, for several extragalactic sources including Mrk 421 were reported
by (Tar$\ddot{a}$sranta et al. 2004, 2005). NIR data for three
decades, for several blazars including Mrk 421, were given by
(Fan \& Lin 1999). A much systematic and comprehensive study of this
source was done by Gupta et al. (2004) in the same period of the campaign
for which the present paper is written. In the compiled optical data for
long term observations, variation of 4.6 mag was reported by Stein et al.
(1976) and rapid variability of 1.4 mag in 2.5 hours was reported by
Xie et al. (1988). There are several simultaneous X-ray and gamma-ray
as well as multi-wavelength campaigns for the source (Makino et al. 1987,
Macomb et al. 1995, Takahashi et al. 2000, Katarzynski et al. 2003,
Blazejowski et al. 2005).

In the present paper, we aimed to search for correlated multi-wavelength
variability in Mrk 421. This kind of study will be an important tool for
understanding the emission mechanism of blazars. This paper is structured
as follows. In section 2, we present multi-wavelength observations and data
reduction; section 3 and 4 give the results and discussions of the present
work.


\section{Observations, Data and the Data Reduction}
\label{sect:Obs}

\subsection{TeV Observations with PACT}

\begin{figure}
   \centering
   \includegraphics[angle=0,width=4in]{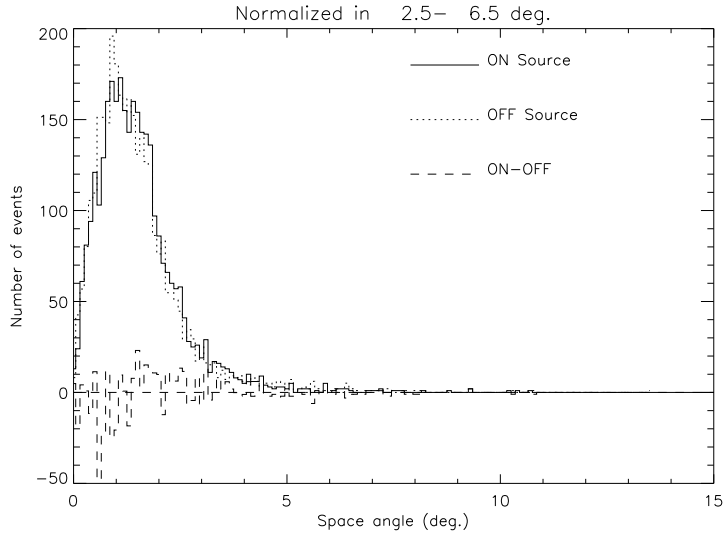}
   \caption{Space-angle distribution of source and background events for
   a typical run.}
\end{figure}

We have used Pachmarhi Array of \v Cerenkov Telescopes (PACT), for observation
of Mrk 421 in TeV gamma rays.
PACT is located in Central India (latitude 22$^\circ$ 28$^\prime$ N, longitude
78$^\circ$ 25$^\prime$ E, altitude 1075 $m$).
We use wavefront sampling technique to
detect {\it TeV} $\gamma$-rays from astronomical sources. There are 24
telescopes spread over an area of $80m \times 100m$. Each telescope has 7
para-axially mounted parabolic mirrors (f/d $\sim$ 1) of diameter 0.9m with a PMT
(EMI 9807B) mounted at the focus of each mirror.
Each telescope is independently steerable, controlled remotely
and monitored throughout the observation (Gothe et al. 2000).
The entire array is sub-divided into 4 sectors with 6 telescopes in each.
Each sector has its own data acquisition system (DAQ) where data on real time,
relative arrival time of PMT pulses (using TDCs) and photon density (using ADCs)
are recorded. A Master DAQ at the center of the array is also used for
recording informations of an event relevant to entire array.
The 7 PMT pulses of a telescope are also linearly added to form a
telescope pulse for trigger generation. Data recording is initiated when a
coincidence of 4 out of any 6 telescope pulses generates an event trigger for a
sector. Typical trigger rate was about 1-3 Hz per sector. The details of this
array are given in (Bhat et al. 2000, Majumdar et al. 2003).

Observations are carried out in ON-OFF mode on clear moonless nights.
In year 2003 from 26th February to 5th March there was a
world-wide multi-wavelength campaign in several wavebands for this source
including PACT.  During these nights 2 sectors out of 4 were aligned along
the source direction and remaining 2 were looking at a background region
simultaneously. The background region is chosen to be a dark region
with the same declination as that of the source but with different RA.
Background region is chosen in such a way that there is substantial overlap
of  zenith angle range between the source and background runs.
The typical run span was about 1-3 hours. The sectors that look at source
and background were interchanged on a daily basis.

A number of preliminary checks were carried out on the data before doing
actual analysis. It was found that data taken on 26th and 27th February,
3rd and 5th March were very bad therefore rejected. Observations taken on
28th February, 1st, 2nd and 4th March are analyzed.
The arrival direction of each shower is determined by reconstructing
shower front using the relative arrival times of \v Cerenkov photons at
various telescopes (or PMTs). \v Cerenkov photon front is then fitted with
a plane, normal to this plane gives the direction of the shower axis. Then,
for each shower or event, this space angle is estimated as an angle between the
direction of shower axis and the source direction. Thus space angles are
obtained for all the events for source as well as background runs.
Space angle distributions of source runs are compared with respective
background runs over the same zenith angle region. Due to some
technical problem same night's source and background runs could not
be compared for these runs. Each source run is compared with previous
nights or next nights background run so that the geometry of the telescope
setup is also same for Source and Background runs. For this comparison,
the shapes of space angle distributions in
2.5$^\circ$ to 6.5$^\circ$ region of source and background were normalized,
as we do not expect any signal beyond 2.5$^\circ$ (Majumdar et al. 2003).
Normalization of distributions corresponding to the background with
that of the source is necessary as these two data sets were taken
at different times. Differences between the number of source and
normalized background events within 2.5$^\circ$ gives the estimate
of $\gamma$-ray events.
Figure 1 shows the space angle distributions for a typical pair
of source and background runs.
Details of analysis procedure are given in (Bose et al. 2005).
During the campaign nights no excess of events
over the background is detected in any of those four nights, implying
$\gamma$-ray flux is close to or below the sensitivity limit of PACT.

\v Cerenkov photon showers initiated by $\gamma$-rays and protons were
simulated  using CORSIKA air shower simulation (Heck et al. 1998) code
to  estimate trigger rate, energy threshold, collection area etc for the PACT setup.
For $\gamma$-rays  incident vertically the energy threshold, defined as the peak of
differential rate curve, is estimated to be 750 GeV and the corresponding
collection area is 1.58$\times$10$^{5}$~m$^{2}$. For Mkn421, which is at an angle
of 20$^{\circ}$ w.r.t. zenith the energy threshold is estimated to be 1.2 TeV
and the collection area as 1.8$\times$10$^{5}$~m$^{2}$.

\subsection{X-Ray Observations with RXTE}

We have analyzed Mrk 421 data observed with RXTE during 26/2/2003 -
6/3/2003. We have extracted archival data sets corresponding to this
multi-wavelength campaign under the guest observing program 80172.
RXTE has two types of detectors viz, Proportional Counter Array (PCA)
and High-Energy X-ray Timing Explorer (HEXTE) on-board along with
All Sky Monitor (ASM). The PCA consists of five identical xenon filled
proportional counter units (PCUs) covering an energy range of 2-60 keV.
During these observations only PCU 0 and PCU 2 were used. Since PCU 0
lost pressure in the top veto at the beginning of Epoch 5,
we have used only data from PCU 2. HEXTE consists of two clusters
of phoswich scintillation detectors covering an energy range of 15-250
keV, but is less sensitive. We do not discuss HEXTE data here. ASM
consists of three xenon filled position sensitive proportional counters
with field of view of 6 $\times$ 90 degrees. It covers 80\% of the sky every
90 minutes and spans an energy range of 2-10 keV.

We analyzed Standard 2 PCA data which has a time resolution of 16s
with energy information in 128 channels. Data reduction is done with
FTOOLS (version 5.3.1) distributed as part of HEASOFT (version 5.3).
For each of the observations, data was filtered using standard procedure
for faint sources given in RXTE Cook Book. For extraction of background,
model appropriate for faint sources (pca\_bkgd\_cmfaintl7\_eMv20031123.mdl)
was used. Light curves were extracted from data for three energy bands:
2-9, 9-20 and 20-40 keV. Background light curves were also extracted
and subtracted from source light curves.
We obtained ASM data from MIT archive. Light curves were generated
taking one-day average.

\subsection{Data from Literature: Multi-wavelength Data}

Near Infrared data in J band used in the present paper is taken from
(Gupta et al. 2004). They have done observations from 1.2 meter
optical/NIR telescope at Gurushikhar observatory, Mount Abu, India
using NICMOS-3 detector and J band filter. The detail about NIR
observations and data reductions is given in (Gupta et al. 2004).

The radio data is taken from the recent paper by (Tar$\ddot{a}$sranta
et al. 2005). They observed the source during the campaign by their
17.7 meter Mets$\ddot{a}$hovi radio telescope at 22 and 37 GHz. The
detail about radio data is given in (Tar$\ddot{a}$sranta et al. 1998).

\begin{figure}[h]
   \centering
\includegraphics[angle=0,width=4in]{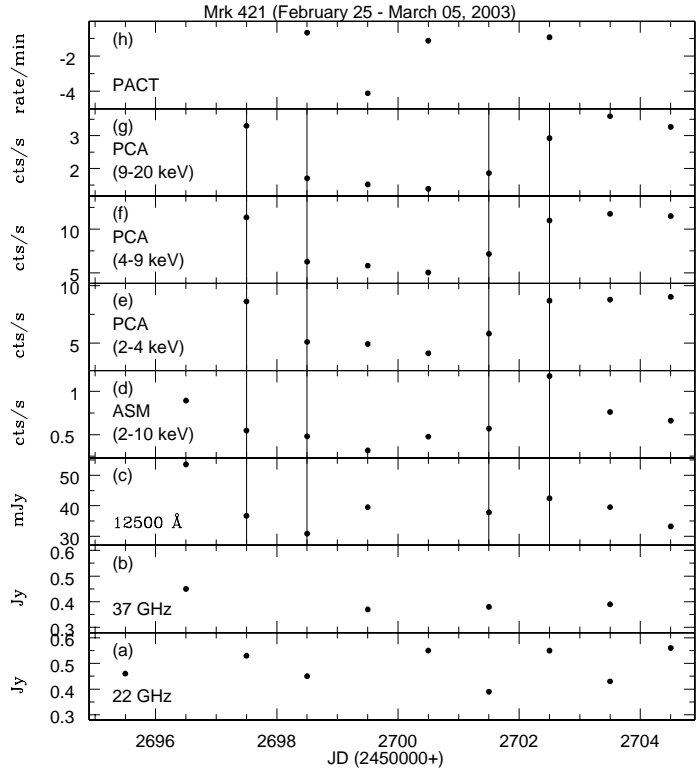}
\caption{Multi-wavelength data of Mrk 421 as a function
of time for all bands from radio to gamma-rays observed during February 25
$-$ March 05, 2003. Vertical lines in panel (c)$-$(g) show simultaneous
variability in NIR and X-ray bands. In general uncertainties are
smaller than the symbols, the error bars have been omitted.}
\label {fig2}
\end{figure}

\section{Results}

\subsection{Multi-wavelength Light Curves}

Figure 2 gives the radio to gamma-ray light curves for the multi-wavelength
campaign during February 25 - March 05, 2003. The data plotted here for
different bands of the EM spectrum is daily average. Daily average of a
specific date is reported at 00h 00m 00s UT. The radio flux seems to be
in stable state, implies variability timescale may be longer than the
duration of the campaign. On the other hand gamma-ray data is noisy.
The figure shows highly correlated variability among the different energy
bands of the PCA data.

\begin{figure}
   \centering
   \includegraphics[angle=0,width=4in]{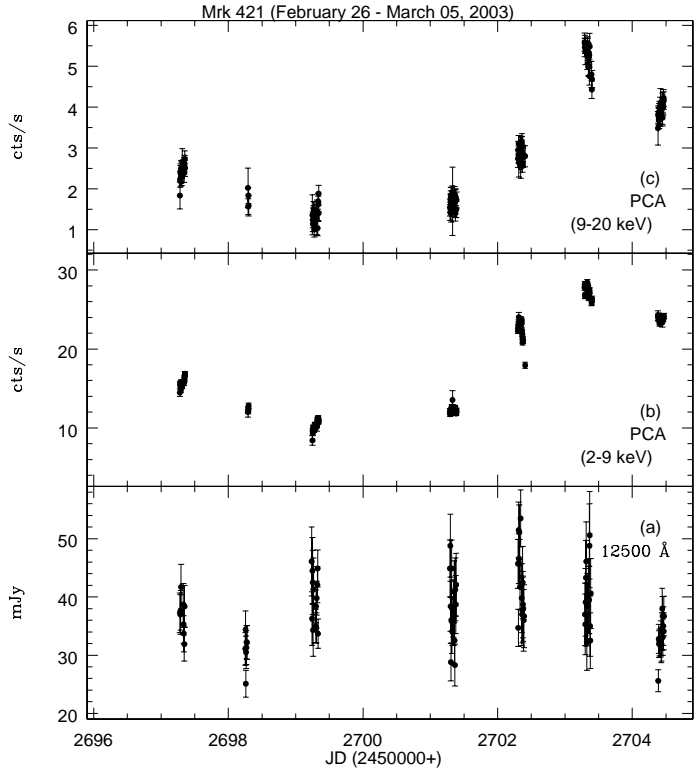}
   \caption{NIR and X-ray bands data of Mrk 421 as a function
of time observed during February 26 $-$ March 05, 2003.}
   \label{fig3}
\end{figure}

Figure 3 gives the NIR and X-ray light curves (with 5 minutes binning)
for the observations during February 26 - March 05, 2003. X-ray
coverage was much longer than the NIR coverage therefore we have selected
that portion of X-ray data which was approximately simultaneous with NIR.
For this plot PCA data of 2 energy bands, 2-4 keV and 4-9 keV are combined.

\subsection{Cross correlation function (ZDCF)}

We computed ZDCF (Z-transformed discrete correlation function)
(Alexander 1997) from light curves in X-ray and NIR bands. The
ZDCF is a method for determining the cross-correlation function
(CCF) of light curves in different energy bands which have non-evenly
sampled data. The ZDCF makes use of the Fisher's z-transform of the
correlation coefficient. Fisher's z-transform of the linear
correlation coefficient, r, is used to estimate the confidence level
of the measured correlation. This method attempts to correct the
biases that affect the original DCF (discrete correlation function)
by using equal population binning. The ZDCF involves the following
three steps:  \\
(i) All possible pairs of observations, (a$_{i}$, b$_{j}$), are
sorted according to their time-lag $t_{i} - t_{j}$, and binned into
equal population bins. \\
(ii) Each bin is assigned its mean time-lag and the intervals above
and below the mean that contain 1$\sigma$ of the each point. \\
(iii) The correlation coefficients of the bins are calculated and
z-transformed. The error is calculated in z-space and transformed back
to r-space.

The time-lag corresponding to the ZDCF is assumed as the time delay
between both components. This function is much more efficient in
detecting any correlation present also it takes care of the data gaps.
The ZDCF seems to peak at a negative lag $-$3 days, which implies that
X-ray variability lags the NIR variability. Since the dataset is 
sparse and the value of ZDCF (max) is $\sim$ 0.5, so, we claim it as
a weak correlated variability. It will be interesting to see with more 
such observations with focused effort in future.    

The ZDCFs are plotted in figure 4 for the three combination of light
curves plotted in figure 3. The cross-correlation coefficient max(ZDCF) and time lag
$\tau$  for each combination of light curves are as follows:  \\
(i) ~~For PCA (2$-$9 keV) vs 12500 $\AA$ \\
\hspace*{0.24in} max(ZDCF) = 0.530$^{+0.1715}_{-0.1515}$, $\tau$ = $-$3.02 days \\
(ii) ~For PCA (9$-$20 keV) vs 12500 $\AA$ \\
\hspace*{0.24in} max(ZDCF) = 0.460$^{+0.1915}_{-0.1728}$, $\tau$ = $-$3.02 days \\
(iii) For PCA (2$-$9 keV) vs PCA (9$-$20 keV) \\
\hspace*{0.24in} max(ZDCF) = 0.747$^{+0.0173}_{-0.0168}$, $\tau$ = $-$0.11 days

\begin{figure}[t]
   \centering
\includegraphics[width=4in,height=4in]{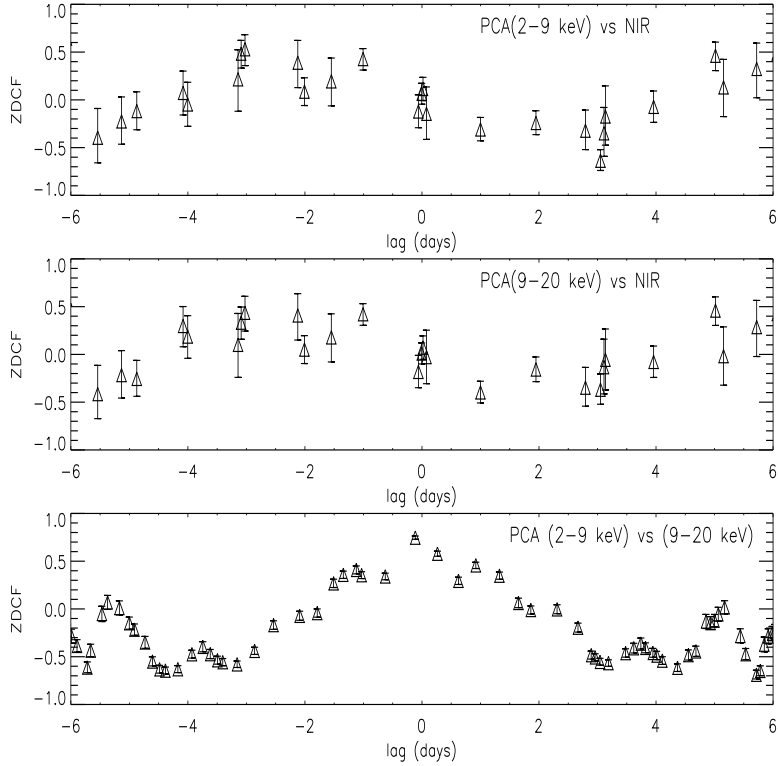}
\caption{NIR and X-ray correlations. Positive lags imply that the second
light curve lags the first.}
\label {fig4}
\end{figure}

The visual inspection of the data in the figure 2 show, a strong correlation 
in NIR and X-ray bands at JD 2452697.5, 2452698.5, 2452701.5 and 2452702.5. 
In particular, the source has tendency to come to faint stage at JD 2452697.5 
and a flaring activity at JD 2452701.5. Thus the positive correlation is seen 
in both in flaring state and quiescent state. At JD 2452699.5, NIR has shown 
anti-correlation with PCA data (this anti-correlation is responsible for lowering 
the correlation coefficients mentioned above). On March 01, 2003 observations 
could not be taken in NIR J band due to bad weather condition. So, NIR data for 
JD 2452700.5 is not present in the panel (c) of the figure 2.

\subsection{Spectral Energy Distribution (SED)}

The SED of Mrk 421 is plotted in the figure 5 in the form log $\nu$F$_{\nu}$
vs. log $\nu$. All frequencies used here are observed frequencies. The
synchrotron component of the SED was fitted using NIR and X-ray data with 
a parabolic function
\begin{eqnarray}
y = Ax^{2} + Bx + C
\end{eqnarray}
the synchrotron peak frequency is determine by $\nu_{peak} = -$B/2A. The
$\nu_{peak} =$ 16.30 is calculated from the figure 5. It has been noticed
for HBLs by Nieppola et al. (2006) that simple parabolic function produces
some error. For HBLs, the synchrotron peak is expected close to soft X-ray
band which decline to be very rapid. Thus in the real sense, the peak
frequency will be slightly higher to the reported value.

\begin{figure}[h]
   \centering
\includegraphics[width=4in,height=4in]{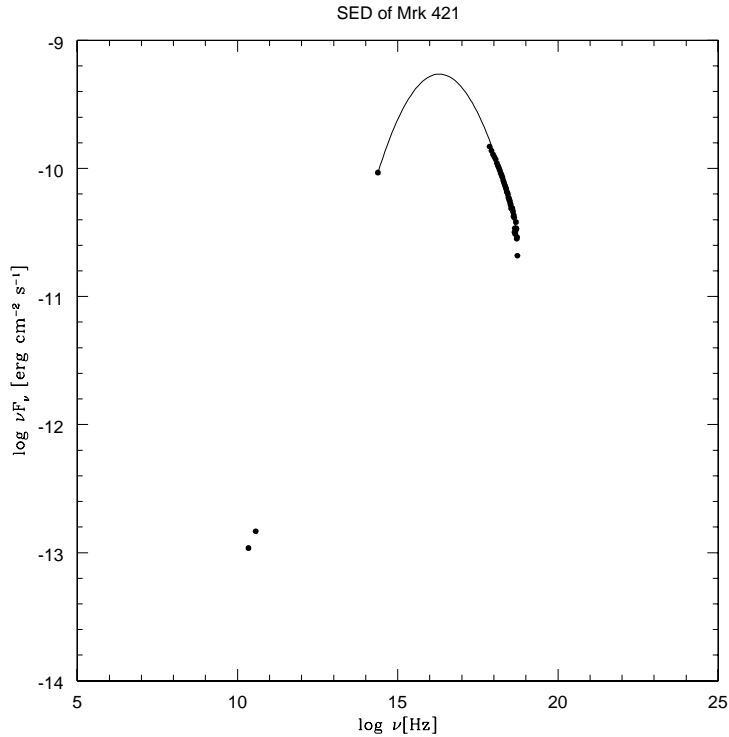}
\caption{Spectral energy distribution of the Mrk 421.}
\label {fig5}
\end{figure}

\section{Discussions}

We have found here the evidence of a possible correlation between X-ray 
and NIR wavelengths in Mrk 421 which is an HBL blazar.

So far, correlated X-ray and NIR variation was observed in a LBL blazar 3C 273,
a flat spectrum radio quasar (FSRQ) (McHardy et al. 1999). For another LBL blazar
AO 0235$+$164, simultaneous radio and optical variability was reported during the
optical outbursts occurred in 1975, 1979 and 1997 (MacLeod et al. 1976; Ledden et
al. 1976; Rieke et al. 1976; Balonek \& Dent 1980; Webb et al. 2000) on the other
hand no correlation was found between radio and optical emission of another LBL
blazar S5 0716+714 (Ostorero et al. 2006). In a multi-wavelength campaign of LBL 
blazar 3C 279 reported by (Wehrle et al. 1998), correlated variability in X-ray 
(PCA) and EGRET CGRO is seen without any time lag; UV leading gamma-ray by $\sim$ 
2.5 days. Since UV and gamma-ray data was less, so its reliability is doubtful. 

Edelson et al. (1995) have reported correlated emission between X-ray, UV and optical 
emissions from PKS 2155-304 (an HBL BL Lac object). In another multi-wavelength campaign 
of PKS 2155-304 from 10 days data, Urry et al. 1997 have reported X-ray flare leading 
EUV flare by one day and UV flare by two days. McComb et al. (1995) and Maraschi et al. 
(1999) have reported simultaneous X-ray and $\gamma-$ray flares in Mrk 421 in two different
campaigns. In other campaign of Mrk 421 in 1998 (Takahashi et al. 2000), complex
variability, positive and negative lags were found which authors report may not be real,
If the lag from both signs are real, these imply that particle acceleration and X-ray cooling
timescales are similar. Katarzynski et al. (2003) show a well defined correlation between 
observed radio outburst in Mrk 421 with a corresponding X-ray outburst and a $\gamma-$ray 
flare in TeV range. In simultaneous TeV and optical observations of blazar 1406$-$076, optical 
flare leading gamma-ray flare by $\sim$ 1 days (Wagner et al. 1995).

In a recent paper, Villata et al. (2006) have shown, in the historical radio and optical
light curves variability behavior is different, while prominent and long-lasting radio
outbursts were visible at various radio frequencies, with higher-frequency variations
preceding the lower-frequency ones. After that date, the optical activity increased
and the radio flux is less variable, suggest that optical and radio emissions come from
two separate and misaligned jet regions. The correlated emissions reported in several papers
here, support models involving single populations of relativistic electrons responsible for
the emission.

One of the important question is ``Where does the NIR radiation originate"?
There are three possibilities: (i) emission from circumstellar dust (ii)
emission from the accretion disk (iii) synchrotron emission by relativistic
electrons in the jet, the first two are external to the jet.
The variability behavior is indicating the similar origin for NIR and
X-ray emission. X-ray emission is originating from synchrotron radiation by
electrons in this AGN as mentioned before.

From different multi-wavelength campaigns it has been noticed that variations on 
longer wavelengths are generally not seen during X-ray and $\gamma-$ray flaring 
events (Tosti et al. 1998 and references therein). Ghisellini \& Maraschi (1996) 
have shown that the overall emission of the BL Lac Mrk 421 could be explained using a
homogeneous SSC model, in which the equilibrium particle distribution is found
balancing continuum injection, cooling and particle escape. In this, single
population of relativistic electrons emit synchrotron radiation up to the UV
or X-ray band and soft photons upto the IR-optical bands undergo upscattering
with the most energetic electrons to form TeV $\gamma-$rays.
However, Blazejowski et al. (2005) have found TeV flares
reached its peak days before X-ray flare during a giant flare or outburst in 2004.
Spectral energy distribution (SED) generated by Blazejowski et al. (2005) was not
fitted with one-zone synchrotron self-Compton (SSC) model but could be fitted
well with additional zones.

Following Ghisellini \& Maraschi (1996), Marscher (1996) and other authors,
we note that the lags in the multi-frequency light curves of Mrk 421 (of
the order of hours between soft and medium X-ray photons, near simultaneous
X-ray and $\gamma-$ray flares, IR leading X-rays by couple of days) require
energy stratification in the source. Similar delays were noticed in 3C 279
(IR leads X-ray by 0.75$\pm$0.25 day) and in PKS 2155$-$304
(a day between EUV and X-rays, 2 days between UV and X-rays) and of
the order of few hours between these (Edelson et al. 1995) as well.
Frequency stratification and different time scales for the duration of these
flares (shorter times for higher frequencies) are possible with the shock-in-jet
model (Marscher 1996 and references therein).

Within the context of the shock-in-jet model of AGNs, we attribute the NIR
emission component to the internal shock driven into the jet by the variation
of the central engine. The correlated X-ray and NIR variability with time
lag of few days is a strong probe of the jet emission not affected much by
the environment of the surrounding medium. The anti-correlation seen in the
NIR and soft X-ray light curve at JD 2452699.5 may be due to the reverse
shock arising by the jet's collision with the surrounding medium.

\begin{acknowledgements}
We thankfully acknowledge the referee for useful comments.
The work of A. C. Gupta and J. H. Fan is supported by the National Natural Scientific 
Foundation of China (grant no. 10573005 and 10125313). We gratefully acknowledge the 
use of RXTE data from the public archive of GSFC/NASA. Also, we are grateful to all 
the members of PACT collaboration for their respective contributions.
\end{acknowledgements}

\end{document}